\documentclass[preprintnumbers, floatfix, 
preprintnumbers, letterpaper, superscriptaddress,nofootinbib, twocolumn]{revtex4}

\usepackage{graphicx}
\usepackage{microtype}
\usepackage{amsmath}
\usepackage{amssymb}
\usepackage{subfigure}
\usepackage{hyperref}
\usepackage{url}
\usepackage{xcolor}
\usepackage{color}
\usepackage{mathrsfs}
\usepackage{calrsfs}
\usepackage{amsfonts}
\usepackage{latexsym}
\usepackage{ragged2e}
\usepackage{epsfig}
\usepackage{textcomp}

\usepackage{caption}
\DeclareCaptionJustification{justified}{\leftskip=0pt \rightskip=0pt \parfillskip=0pt plus 1fil}
\definecolor{vividviolet}{rgb}{0.62, 0.0, 1.0}
\definecolor{amaranth}{rgb}{0.9, 0.17, 0.31}
\definecolor{palatinateblue}{rgb}{0.15, 0.23, 0.89}
\definecolor{brightpink}{rgb}{1.0, 0.0, 0.5}
\definecolor{cornflowerblue}{rgb}{0.39, 0.58, 0.93}
\definecolor{deepcarminepink}{rgb}{0.94, 0.19, 0.22}
\definecolor{radicalred}{rgb}{1.0, 0.21, 0.37}

\hypersetup{ linktoc=all,
    colorlinks, linkcolor={palatinateblue},
    citecolor={brightpink}, urlcolor={amaranth}
}
\newcommand{\changeurlcolor}[1]{\hypersetup{urlcolor=#1}}
\graphicspath{{Images/}}

\renewcommand{\d}[1]{\ensuremath{\operatorname{d}\!{#1}}}

\newcommand{\be}{\begin{equation}}
\newcommand{\ee}{\end{equation}}
\newcommand{\bs}{\begin{split}} 
\newcommand{\bea}{\begin{eqnarray}}
\newcommand{\eea}{\end{eqnarray}}

\def\sideremark#1{\ifvmode\leavevmode\fi\vadjust{\vbox to0pt{\vss% the remark
 \hbox to 0pt{\hskip\hsize\hskip1em%                          will appear only
 \vbox{\hsize1.2cm\tiny\raggedright\pretolerance10000%          on the side
 \noindent #1\hfill}\hss}\vbox to8pt{\vfil}\vss}}}%
\begin{document}

\title{The Quantum Atmosphere of Reissner-Nordstr\"om Black Holes}

\author{Yen Chin \surname{Ong}}
\email{ycong@yzu.edu.cn}
\affiliation{Center for Gravitation and Cosmology, College of Physical Science and Technology, Yangzhou University,\\ Yangzhou 225009, China}

\author{Michael R. R. \surname{Good}}
\email{michael.good@nu.edu.kz}
\affiliation{Department of Physics \& Energetic Cosmos Laboratory, Nazarbayev University, \\Nur-Sultan 010000, Kazakhstan}

\begin{abstract}
Hawking radiation originates from a ``quantum atmosphere'' around black holes,  not necessarily from the vicinity of the horizon. We examine and discuss the properties of quantum atmospheres of asymptotically flat Reissner-Nordstr\"om black holes, which extends further and further away from the black hole as extremality is approached, though arguably it becomes indistinguishable from normal vacuum fluctuation at spatial infinity. In addition, following our previous findings on re-writing the Hawking temperature of a Kerr black hole in terms of a ``spring constant'', we generalize the same notion to the Reissner-Nordstr\"om case, which allows us to put a minimum size on the location where Hawking particles can be emitted near a black hole, which agrees with the stretched horizon.
\end{abstract}

\maketitle

\section{Introduction: Where Does Hawking Radiation Originate From?}

A popular cartoon picture of Hawking radiation often depicts the Hawking particle pairs as being produced near the vicinity of the black hole horizon. 
Unfortunately, this misunderstanding is widespread even in the literature. The correct picture is that the uncertainty in the position where Hawking particles are created is rather huge.
As shown by Giddings in \cite{1511.08221}, 
for a Schwarzschild black hole in (3+1)-dimensions, the Hawking radiation originates from a ``quantum atmosphere'' that extends some $\mathcal{O}(r_{h})$ away from the horizon at $r_{h}$. 

There are at least two ways to see why the quantum atmosphere extends some distance away from the black hole.
Giddings calculated the wavelength of a typical Hawking quantum (in Planck units $G=\hbar=c=k_B=1$): 
\begin{equation}\label{w}
\lambda_{T_H} = \frac{2\pi}{T_H} = 16 \pi^2 M \approx 79 r_h,
\end{equation}
 where $T_H$ denotes the Hawking temperature. 
He remarked that: ``thus the horizon size is smaller than the thermal wavelength, in contrast to typical discussions of black body radiation.'' (Indeed, for a ball of radiation, which is close to forming a black hole, one finds that $\lambda$ will scale like $\sqrt{M}$ instead of $M$, see Appendix \ref{ball}.)

Heuristically we can interpret this wavelength as the de Broglie wavelength of the photon $\lambda_\text{dB} = 2\pi / E$. That is to say, a Hawking particle has some probability to be created in a sphere with radius $\lambda_{T_H}$, which is about 80 times the Schwarzschild radius. This crude -- but straightforward -- method allows us to appreciate why a typical Hawking particle should \emph{not} be thought of as coming from the vicinity of the horizon. (One might wonder if looking at the wavelength that corresponds to $T_H$ is the right thing to do, since $T_H$ is the temperature at infinity. However, for the purpose of this discussion, the distinction between $T_H$ and the local Tolman temperature is surprisingly rather small, see Appendix \ref{nearhor}.) 

The second method, also pointed out by Giddings, is to consider the geometric optics approximation, in which one finds that the black hole emits radiation with effective area larger than its event horizon. The effective radius that goes into the Stefan-Boltzmann Law is $r_a = 3\sqrt{3} M = (3\sqrt{3}/2)r_h$, which corresponds to the maximum impact parameter for an infalling massless particle to fall onto the photon orbit at $r_{\text{ph}}=3M$ (hence also associated with the potential that an \emph{escaping} massless particle needs to overcome in order to actually escape to null infinity). Therefore we can say that the quantum atmosphere has radius $r_a\sim \mathcal{O}(r_h)$. (See also \cite{unruh}.) The interpretation here is somewhat different however: the effective potential essentially screens the escaping particles so that only sufficiently energetic ones can escape to infinity, it does \emph{not} say anything about \emph{where} the particle was first created. Nevertheless, if only sufficiently energetic ones escape, we can use this idea to carry out a crude statistical estimate of the field solutions for entropy.

We note that $\lambda_{T_H}$ and $r_a$ is not the same, the former is of order $\mathcal{O}(80 r_{h})$ but the latter is only of $\mathcal{O}(r_{h})$. (Note that these are of course, coordinate distances, not physical distances.)
That is, the two definitions of the size of the quantum atmosphere do not agree quantitatively. 
However, \emph{qualitatively}, the main message is the same: the quantum atmosphere of a Schwarzschild black hole extends some distance away from the black hole. Since both methods are rather crudely defined anyway, they should be seen as only approximating the ``true'' extend of the quantum atmosphere (if one has an improved, more precise definition for it).

In \cite{1701.06161}, Dey, Liberati and Pranzetti examined the (semi-classical) stress energy tensor and found that indeed the energy density and fluxes of particles peaked at some radius $\mathcal{O}(r_h)$ from the horizon. They also supplemented the argument with a heuristic one in which the Hawking pairs are separated due to gravitational analogue of Schwinger process (production of charged particle from vacuum, due to strong external electric field \cite{Schwinger}), which can be considered as an improved version of the usual cartoon picture. More recently, together with Mirzaiyan, they have also re-examined the issue from the point of view of a freely falling observer \cite{1906.02958} and found similar results continue to hold. See also \cite{1908.03374} for the case of dimensionally reduced Schwarzschild black hole.

In this work we wish to study the quantum atmosphere for asymptotical flat Reissner-Nordstr\"om black holes. 
(We will work in the units such that the vacuum permittivity satisfies $4\pi \epsilon_0=1$.)
Unlike the Schwarzschild case, the temperature of a Reissner-Nordstr\"om black hole tends to zero in the extremal limit, which means the wavelength $\lambda_{T_H} = 2\pi/T_H$ will diverge in this limit. On the other hand,  the impact parameter $r_a$, which in Reissner-Nordstr\"om case takes the form
\begin{equation}
r_a=\frac{1}{2\sqrt{2}}\frac{(3M+\sqrt{9M^2-8Q^2})^2}{\sqrt{3M^2-2Q^2+M\sqrt{9M^2-8Q^2}}},
\end{equation}
tends to $4M$ in the extremal limit $M \to Q$. This means that the difference between $\lambda$ and $r_a$ can be very large, so these two quantities do not generally agree even qualitatively. Therefore the study of the quantum atmosphere of Reissner-Nordstr\"om black holes is well motivated. 

We shall see in Sec.(\ref{heuristic}) that the heuristic ``gravitational Schwinger effect'' argument of \cite{1701.06161} does support the idea that the quantum atmosphere scales as the wavelength of the typical Hawking quanta, instead of the impact parameter of the photon orbit. We shall give further argument for this in the Discussion, by considering asymptotically locally anti-de Sitter (AdS) black holes. In addition, in Sec.(\ref{small}) we shall study, in the asymptotically flat case, just how close can Hawking quanta emerge from the vicinity of a black hole, by taking into account the entropy content in the spherical shell around the black hole within its photon orbit, as performed in Sec.(\ref{states}). Part of the calculations is facilitated by writing the Hawking temperature in terms of the ``spring constant'', first introduced in the context of asymptotically flat Kerr black holes \cite{1412.5432}.
 
\section{Heuristic Argument for the Quantum Atmosphere of Reissner-Nordstr\"om Black Holes}\label{heuristic}

Following \cite{1701.06161}, we first calculate the tidal acceleration at some coordinate distance $r=r_*$ from the black hole. For a general spherical symmetric black hole with metric function $g_{tt}=-f(r)=-g^{-1}(r)$, the radial tidal acceleration is given by \cite{1602.07232}
\begin{equation}
a^r|_{r_*}=-\frac{f''}{2}n^r,
\end{equation} 
so for Reissner-Nordstr\"om black hole we obtained
\begin{equation}
a^r|_{r^*}=\left(\frac{2M}{r_*^{3}}-\frac{3Q^2}{r_*^4}\right)n^r.
\end{equation}
We have, again following \cite{1701.06161}, the approximation $n^r \sim \lambda_c =\hbar/mc=1/m$ in our units.

The radial component of the free fall velocity of the outgoing particle is
\begin{equation}
u^r=\frac{\d r}{\d \tau}=\sqrt{E^2-1+\frac{2M}{r_*}-\frac{Q^2}{r_*^2}},
\end{equation}
where $E$ is the energy of the particle at infinity. We can choose it to be unity (in the notation of \cite{1701.06161}, this is equivalent to setting $r_0=0$ therein),
so that 
\begin{equation}
u^r=\frac{\d r}{\d \tau}=\sqrt{\frac{2M}{r_*}-\frac{Q^2}{r_*^2}}.
\end{equation}

In the static observer's frame, we have %\edz{Ref.\cite{1701.06161} has another mistake here, the power should be -1/2 not -1.}
\begin{equation}
a^r_{\text{st}}=a^r \cosh(\zeta)=a^r \left(1-\frac{2M}{r_*}+\frac{Q^2}{r_*^2}\right)^{-\frac{1}{2}},
\end{equation}
where $\zeta=\tanh^{-1}(u^r)$ is the rapidity.

The radial component of the force under this transformation is given by the relativistic Newton's second law 
\begin{equation}
F^r_\text{tidal-st} =  \left.\frac{ma^r_{\text{st}}}{\sqrt{1-\frac{2M}{r}+\frac{Q^2}{r^2}}^3}\right\vert_{r_*}= \left.\frac{ma^r}{\left(1-\frac{2M}{r}+\frac{Q^2}{r^2}\right)^2}\right\vert_{r_*},
\end{equation}
which yields
\begin{equation}
F^r_\text{tidal-st} =  \frac{m}{\left(1-\frac{2M}{r_*}+\frac{Q^2}{r_*^2}\right)^2}\left(\frac{2M}{{r_*}^3}-\frac{3Q^2}{r_*^4}\right)\lambda_c,
\end{equation}
at which point $m$ cancels with $\lambda \sim 1/m$.

The magnitude of the radial tidal force is thus
\begin{flalign}
\left\| F^r_\text{tidal-st} \right\| &=\sqrt{g_{rr}F^r_\text{tidal-st} F^r_\text{tidal-st} }  \notag \\
&=\left(\frac{2M}{{r_*}^3}-\frac{3Q^2}{{r_*}^4}\right)\left(1-\frac{2M}{r_*}+\frac{Q^2}{{r_*}^2}\right)^{-\frac{5}{2}}.
\end{flalign}

The work required by the tidal force to split the particle pair apart is thus
\begin{equation}
W_{\text{tidal}} \sim \left\| F^r_\text{tidal-st} \right\| d(r_*)
\end{equation}
where
\begin{equation}
d(r_*)=\int_{r_h}^{r_*} \sqrt{g_{r'r'}} \d r',
\end{equation}
with $r_h=M+\sqrt{M^2-Q^2}$ being the outer (event) horizon of the black hole. 

\begin{figure}[!h]
\centering
\includegraphics[width=3.4in]{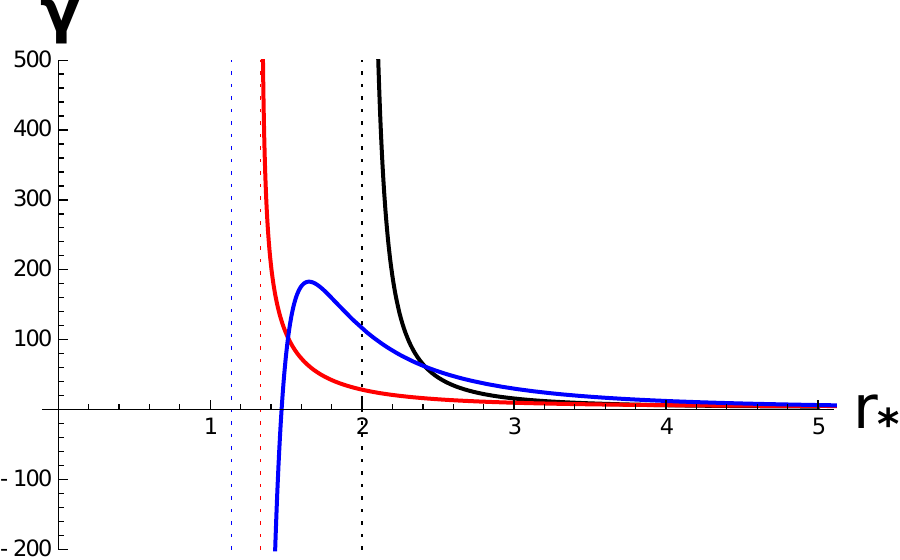}
\caption{The plot of $\gamma$ as function of areal radius. We set $M=1$. The right most curve (black) corresponds to the Schwarzschild $(Q=0)$ case, the other curve that diverges to $+\infty$ as $r \to r_h$ (red) corresponds to the special value $Q=2\sqrt{2}/3$ beyond which the curve would develop a global maximum and turns around: an example is provided with the remaining curve (blue), with $Q=0.99$. Dotted lines indicate the event horizons for each case. The blue curve eventually tends to the horizon as well, but not shown at this scale.}
\label{gamma}
\end{figure}

The frequency of a typical Hawking particle at infinity is (we momentarily restore $k_B$ and $\hbar$ for clarity):
\begin{equation}
\omega_{\infty} =\frac{\gamma}{\hbar} k_B T_H = \frac{\gamma k_B}{\hbar}\left[\frac{1}{2\pi}\frac{\sqrt{M^2-Q^2}}{(M+\sqrt{M^2-Q^2})^2}\right].
\end{equation}
Therefore at distance $r=r_*$, we have $\omega_{r_*}=\omega_\infty/\sqrt{g_{00}}$. We can now solve for $\gamma=\gamma(r_*)$ via $W_{\text{tidal}}=2\omega_{r_*}$, i.e. we equate the work to the total energy of the two Hawking quanta being created. 
Thus, we obtain
\begin{flalign}
\gamma(r_*)=& \pi \left(\frac{2M}{r_*^3}-\frac{3Q^2}{r_*^4}\right)\left(1-\frac{2M}{r_*}+\frac{Q^2}{r_*^2}\right)^{-2} \notag \\
&\cdot\frac{(M+\sqrt{M^2-Q^2})^2}{\sqrt{M^2-Q^2}}d(r_*).
\end{flalign}
This can be plotted numerically, see Fig.(\ref{gamma}). 
%\begin{figure}[!h]
%\centering
%\includegraphics[width=3.4in]{gamma}
%\caption{The plot of $\gamma$ as function of areal radius. We set $M=1$. The right most curve (black) corresponds to the Schwarzschild (Q=0) case, the other curve that diverges to $+\infty$ as $r \to r_h$ (red) corresponds to the special value $Q/M=2\sqrt{2}/3$ beyond which the curve would develop a global maximum and turns around, an example is provided with the remaining curve (blue), with $Q=0.99$. 
%\label{gamma}}
%\end{figure}

It can be shown that for charge-to-mass ratio $Q/M \leqslant 2\sqrt{2}/3$, the curve is monotonically increasing as we decrease the radius, and would in fact diverge as $r_* \to r_h$. However, if $Q/M > 2\sqrt{2}/3$, then the curve would initially increase as we decrease $r_*$, however, it eventually turns around and goes to zero at some point, so that the function $\gamma$ is negative near the horizon (in fact diverges to $-\infty$ as one tends to the horizon).  To see this, one simplify verifies that $\gamma$ has a zero at $r_*=(3/2)(Q^2/M)$, which is only real if $r_*>r_h$, i.e. $Q/M > 2\sqrt{2}/3$. Increasing the charge further would raise the value of the global maximum of the curve.

According to \cite{1701.06161}, solving the equation  $\gamma(r_*)=2.82$ would then yields the location for the quantum atmosphere, where $3+ W\left(-3e^{-3}\right) = 2.82$ being the famous number that appears in the Wien's displacement law for thermal radiation $h\nu_\text{max}=2.82~kT$.
However, there are some complications here for the charged case.
Clearly, for $Q/M > 2\sqrt{2}/3$, there are two solutions for the equation $\gamma(r_*)=2.82$, one of which is near horizon and the other one becomes further and further away as $r_* \to \infty$. We have explicitly (with $M=1$), $r_*(Q=0)=5.2592$, $r_*(Q=2\sqrt{2}/3)=5.3979$, whereas $Q=0.99$ gives $r_*=6.7737$ or $r_*=1.4710$. 
Naively, this means that there are \emph{two} locations $r_{*1,2}$ where most of the radiation is created: one of them, $r_{*1}$, remains close to the horizon while the other one, $r_{*2}$, is moving outward as charge-to-mass ratio increases, eventually diverges to infinity in the extremal limit. Thus $r_{*1}$ is qualitatively the same as the behavior for the effective emission radius $r_a$, while $r_{*2}$ behaves like the wavelength $\lambda_{T_H}$ (though here $r_{*2}$ corresponds to the radius of a spherical shell which has the peak in the emission, whereas the wavelength does not really tell us where the peak is, it only gives the natural scale involved).

\begin{figure}[!h]
\centering
\includegraphics[width=3.4in]{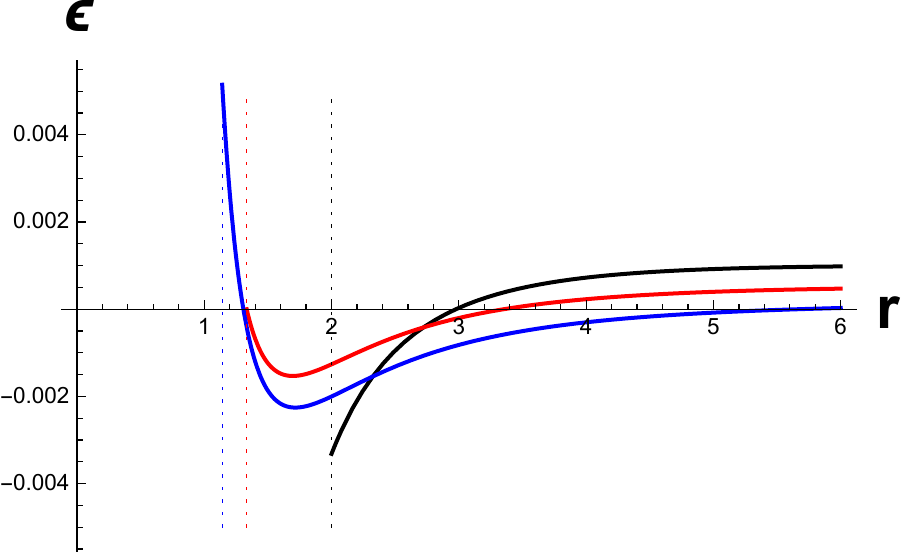}
\caption{The plot of $\epsilon$ as function of areal radius, plotted in the domain $r\in[r_h, 5]$. We set $M=1$. The black curve, which is monotonically increasing in $r$, corresponds to the $Q=0$ case, is negative near the horizon. For  $Q=2\sqrt{2}/3$, which corresponds to the red curve, $\epsilon$ tends to zero as $r \to r_h=4/3$. For $Q$ larger than $2\sqrt{2}/3$, for example, $Q=0.99$, which is depicted as the blue curve, $\epsilon$ becomes positive around the horizon. Dotted lines indicate the event horizons for each case\label{epsilon}.}
\end{figure}

The fact that $\gamma(r_*)$ becomes negative near the horizon for $Q/M > 2\sqrt{2}/3$ mirrors the behavior of the expectation of time-time component of the stress energy tensor, $\left\langle T_t^{~t}\right\rangle$.
Following  Loranz and Hiscock \cite{9607048}, we define ``energy density'' 
\begin{equation}
\epsilon:=-\left\langle T_t^{~t} (r)\right\rangle = \frac{f''(r)}{24\pi} - \frac{[f'(r)]^2}{96\pi f(r)} + \frac{\pi T_H^2}{6 f(r)},
\end{equation}
which becomes \emph{positive} near the horizon under the \emph{exact same condition} that  $Q/M > 2\sqrt{2}/3$ , see Fig.(\ref{epsilon}).

Thus, despite the heuristic treatment above, the change of the behavior of particle production near extremality thus discovered agrees with that obtained from quantum field theoretic calculation of $\epsilon$.
If we define the ``quantum atmosphere'' as the largest areal radius at which the energy density becomes positive (such a choice of $r_{*2}$ over $r_{*1}$ is preferred by continuity; also see later discussion), then the quantum atmosphere of a near extremal black hole can be very large indeed. (Though recall that, even for $Q/M=0.99$, the value of the areal radius is still  reasonably small: $r_*= 6.7737)$. 

Of course $r_*$ eventually diverges when $Q \to M$ in the extremal limit. Nevertheless, in that case the temperature of the black hole is approaching zero, so the actually probability of a Hawking particle emerging out of the vacuum far away from the black hole is negligibly small and arguably cannot be distinguished from vacuum pair production due to fluctuation in the absence of a black hole anyway. That is, although the quantum atmosphere tells us \emph{where} we can expect Hawking particle to be emitted from, it does not tell us how \emph{frequent} such emission is to be expected.

We mentioned that the reason to prefer $r_{*2}$ over $r_{*1}$ as the definition for quantum atmosphere is due to continuity, that is to say, because $r_{*1}$ only appears once the charge is high enough: $Q/M> 2\sqrt{2}/3 $. This is when the curve $\gamma$ develops a global maximum and turns around, and also when the energy density $\epsilon$ becomes positive near the horizon. In addition, remarkably, as shown by Brynjolfsson and Thorlacius, $Q/M> 2\sqrt{2}/3$ is also when freely falling observer would not detect any radiation near the black hole \cite{0805.1876}. Specifically, said observer would not detect radiation if \cite{0805.1876}
\begin{equation}
r < \frac{2Q^2}{(M^2-Q^2)^{\frac{1}{4}}\left[(M^2-Q^2)^{\frac{1}{4}}+\sqrt{4M-3\sqrt{M^2-Q^2}}\right]}.
\end{equation}
One can verify numerically that this encompasses $r_{*1}$. Indeed the right hand side expression in the inequality above is increasing with $Q$ (for fixed $M$), so that as one increases the charge from $Q>(2\sqrt{2}/3)M$ to $M$, the radius within which a freely falling observer does not see radiation will increase outward from $r=4M/3$ to infinity, whereas $r_{*1}$ is decreasing towards the horizon.

\section{Entropy by State Counting Inside Photon Sphere}\label{states}

The modes inside a large scalar atmosphere can be used to intuitively think about the entropy of a black hole as its area from an underlying statistical counting argument. 
In a $(3+1)$-dimensional spacetime, the number of microscopic states of a massless scalar field living in the metric of a Schwarzschild black hole must scale as the frequency cubed, $\omega^3$, of the field, 
\be N(\omega) \sim \omega ^3.\ee
For clarity, we re-instate dimensionful constants to ensure correct units.  Note that $[\hbar G] = L^5 T^{-3}$. We write
\be N(\omega) \sim \frac{\omega^3\lambda^5}{\hbar G},\ee
where $\lambda$ is the scalar atmosphere diameter surrounding the black hole, equal to twice the well-known photon sphere radius, $\lambda = 2r_p = 3r_h = 6GM/c^2$. 

From Sec.~(\ref{heuristic}), the quantum atmosphere is at $r_* = 5 . 2592$ ($M=1$), which is slightly outside of the photon orbit. Indeed for the charged case, the difference between the two radii gets larger with larger value of the charge. Nevertheless,
our purpose in this section is to compute the entropy within the spherical shell defined by the photon orbit of the neutral case, which will be useful for our later exploration. 

Wavelengths greater than $\lambda$ are more likely to pass over the potential barrier at the photon orbit, while wavelengths smaller than $\lambda$ stabilize inside.
In thermal equilibrium, the particle spectrum described by a Bose-Einstein distribution, has a free energy as a function of temperature with the sum over all energies $\d\omega$ in the available states,
\be F(T) = -\int_0^{\infty}\; \frac{\hbar N(\omega)}{e^{\hbar \omega/k T}-1} \d\omega. \ee
The number of states substituted, gives simply
\be F(T) =-\frac{\lambda^5}{ G } \int_0^{\infty}\; \frac{ \omega^3}{e^{\hbar \omega/k T}-1} \d\omega= -\frac{\lambda^5}{ G } \left(\frac{\pi^4 k^4 T^4}{15\hbar^4}\right).\ee
%\frac{512 \pi^4 G^4 M^5 k^4 T^4}{\hbar^4 c^{10}}.\ee
The entropy is found via $S = -\partial_T F$, which gives
\be S = \frac{4 \pi ^4  \lambda^5 k^4 T^3}{15 G \hbar ^4}.\ee
Using the Hawking temperature, $T = \displaystyle \frac{\hbar c^3}{8\pi G M k} = \frac{c \hbar }{4 \pi  k r_h}$, the entropy is 
\be S = \frac{\lambda ^5}{240 r_h^3}\frac{\pi k c^3}{\hbar G}. \ee
For a sphere with diameter $\lambda = 3r_h$, so that $\lambda^5 = 243 r_h^5$, one obtains:
\be S = \frac{243}{240}\frac{ k c^3}{\hbar G} \pi r_h^2, \ee
or expressed in terms of mass, radius, and area, respectively: 
\bea S &=& \frac{G k}{\hbar c}4\pi M^2 \left(1 + \frac{1}{80}\right),\\
&=&\frac{k c^3}{\hbar G} \pi r_h^2\left(1 + \frac{1}{80}\right),\\
&=&  \frac{k}{\ell_P^2}\frac{A}{4}\left(1 + \frac{1}{80}\right).\eea
This is $S/S_H = 1.0125$ or $1.3\%$ relative error.

In other words, the entropy contained in the sphere of diameter $3r_h$ is a good approximation for the statistical origin of the entropy of the black hole.  As a general heuristic, this underscores the inclusion of an atmospheric contribution to the entropy of the black hole in the form of field mode solutions, if these degrees of freedom are to be counted.

\section{Minimum Position for Hawking Quanta Creation}\label{small}
Despite the large quantum atmosphere, this is not to say that Hawking particle cannot be emitted close to the black hole. 
So how close can virtual particle be created from the vacuum near a black hole?
This measure of closeness can be obtained for the charged black hole by examining the thickness of the brick wall of 't Hooft \cite{tHooft:1984kcu}. 

Here we will make use of the spherical symmetry offered by the RN metric (and switching back to Planck units), \be \d s^2 = -f(r)\d t^2 + f(r)^{-1}\d r^2 + r^2\d\Omega^2,\ee
where our $f(r) \equiv f_q$ is given by
\be f_q: = 1-\frac{r_s}{r} + \frac{r_q^2}{r^2} .\ee
In our units, there are two length scales, $r_q = Q$ (assume positive for simplicity), and $r_s = 2M$.  The outer horizon is located closer to the coordinate origin than $r=r_s$,  and we denote $r_h \equiv r_p$ for positive sign of the square root, 
\be r_p = \frac{1}{2}\left(r_s + \sqrt{r_s^2 -4 r_q^2}\right).\ee
We can perform a series approximation of the radial integrand, of the usual brick wall calculation,
\be \frac{r^2}{f_q^2} \approx \frac{r_p^6}{(r-r_p)^2(2r_p-r_s)^2},\ee
where we have kept only the leading order term around the horizon $r=r_p$ and expressed $r_q$ in terms of $r_p$.  Integrating this from the brick wall position outward gives
\be 
\int_{r_p + b_q}^{\infty}  \frac{r_p^6 }{(r-r_p)^2(r_s-2r_p)^2} \d r= \frac{r_p^6}{b_q \left(2 r_p-r_s\right)^2}.\ee
Since the rest of the calculation is the same, we can identify a new $N_{0}$, defined by:
\be N_q(\omega) \equiv N_{0} \omega^3 := \frac{2}{3\pi}\omega^3 \left(\frac{r_p^6}{b_q \left(2 r_p-r_s\right){}^2}\right).\label{NRN}\ee
The entropy calculation is as before, where
\be S = N_0 \frac{4\pi^4}{15} T^3,\ee
but now we substitute in the colder Hawking temperature of the charged black hole, 
\be T_q = \frac{2r_p-r_s}{4\pi r_p^2} = \frac{1}{4\pi r_s} - \frac{r_s}{4\pi}\left(\frac{1}{r_p}-\frac{1}{r_s}\right)^2.\ee
In the expression on the right, we have separated out the gravitational contribution due to the electric charge (see Appendix \ref{RNtemp}), for clarity.  This gives our $S_q(N_{0})$,  
\be S_q = \frac{\pi  N_{0} \left(2 r_p-r_s\right){}^3}{240 r_p^6}, \ee
which we know must equal the known answer of $S_q = A/4=\pi r_p^2$. Solving for $N_0$ gives,
\be N_0 = \frac{240 r_p^8}{\left(2 r_p-r_s\right){}^3}. \ee
Plugging this into our $N(\omega)$ value, Eq.~(\ref{NRN}), and solving for $b_q$ gives:
\be b_q = \frac{2 r_p-r_s}{360 \pi  r_p^2}.\ee
We can see that as $r_p \rightarrow r_s$, then the usual uncharged result is obtained. It is easy to see that this brick wall is the usual ratio associated with the new temperature, that is, for spherical symmetry,
\be b_q = \frac{T_q}{90}, \ee
which is as it was found in the neutral charge case, $b_s = T_s/90$. This means,
\be  b_q = b_s - \frac{1}{90}\frac{k_q}{2\pi},\ee
where we have underscored the shrinking  contribution due to charge by introducing $k_q$, which is the negative gravitational charge contribution counterpart to the Schwarzschild surface gravity.  As we have already emphasized, this charged black hole calculation is restricted to the spherical symmetry that is left uncorrupted by the addition of charge.  The brick wall acts as a guidepost to just how close particles can be produced at the horizon before complications due to gravitational interactions with the field require quantum gravity. 

Exactly how thick is the brick?  We find this by computing the proper length, which can be calculated in the same way as the uncharged case.  Writing the integral as
\be b_P := \int_{r_p}^{r_p+b_q} \frac{\d r}{\sqrt{f}} = \int_{r_p}^{r_p+b_q} \frac{r \d r}{\sqrt{(r-r_p)(r+r_p-r_s)}},\ee
gives, after imposing the conditions that $2r_p > r_s >r_p>0$ and $b>0$,
\be b_P = \sqrt{b_q (b_q+2 r_p-r_s)}+r_s \sinh ^{-1}\left(\sqrt{\frac{b_q}{2 r_p-r_s}}\right).\ee
For thin bricks, $b_q\ll r_p$, we have, to leading order,
\be b_P = \frac{2 r_p \sqrt{b_q} }{\sqrt{2 r_p-r_s}}.\ee
Plugging in our brick wall, we find
\be b_P = \frac{1}{3\sqrt{10\pi}} \frac{r_p}{r_s}.\ee
This quantity -- which is 0.0595 for the uncharged case -- decreases as more charge is added, until the extremal state is reached: $Q\rightarrow M$, where $r_p \rightarrow r_s/2$, and $b_P^{-1} \rightarrow 6\sqrt{10\pi}$, which gives $b_P = 0.0297$,
about two orders of magnitude smaller than the Planck length. For all ``practical'' purposes, one could treat this as the stretched horizon. In fact, we can treat this minimum size as a property of the horizon. Indeed, similar calculation can be performed for other black holes as well.

\section{Discussion}

In this work, we have investigated the quantum atmosphere of asymptotically flat Reissner-Nordstr\"om black hole. The heuristic ``gravitational Schwinger effect'' argument gives a result that is in exact agreement with the field theoretical calculation of a suitably defined stress energy tensor $\epsilon$. Namely, if we define the largest coordinate radius $r=r_*$ such that $\epsilon(r_*)=0$ and $\epsilon(r>r_*)>0$ to be the quantum atmosphere, at which the Hawking quanta produced will dominate the spectrum, then this atmosphere becomes larger as $Q$ increases. In fact, $r_* \to \infty$ as $Q \to M$ in the extremal limit. Nevertheless, the temperature is also decreasing to zero in the same limit, so pair production rate becomes smaller. Thus, even if the atmosphere goes all the way to spatial infinity, particle production rate is so small that it is arguably indistinguishable form having no black hole. This is exactly what one expects, infinitely far away from the black hole.

The quantum atmosphere defined in this way agrees qualitatively with the proposal that the quantum atmosphere should be proportional to the characteristic wavelength of the typical Hawking quanta, instead of the proposal that it be related to the impact parameter of the photon orbit. This definition for the quantum atmosphere is likely also be helpful in the asymptotically locally AdS cases. As is well-known, there are topological black holes in AdS with either hyperbolic or flat (toral or planar) horizon topology, in addition to spherical ones. Unlike asymptotically flat black holes, the Hawking temperature for AdS black holes is proportional to its size when the black hole is sufficiently large \cite{9808032}. That is, the associated wavelength $\lambda_{T_H}$ is \emph{inversely} proportional to its size, i.e. a sufficiently large black hole can have $\lambda_{T_H}$ that is much smaller than its horizon scale, much like a conventional hot body, which likely means that the Hawking quanta are mostly created close to the horizon (this is consistent with \cite{0805.1876}, in which it was shown that freely falling observer only detects radiation from the black hole when sufficiently close to the horizon). On the other hand, the effective emitting surface -- which in asymptotically flat case corresponds to the photon orbit impact parameter -- has area proportional to $L^2$ where $L$ denotes the asymptotic curvature of AdS \cite{9803061}. For fixed $L$, a large enough black hole will have a horizon $r_h > L$, so that it would not make sense to take $L$ -- now entirely inside the black hole -- as the definition of the quantum atmosphere. 

For the asymptotically flat case, however, the photon orbit still plays an important role. The effective potential associated with the photon orbit traps various modes of the Hawking quanta so that only some with sufficiently large energy can escape. In this sense the effective emitting surface in the geometric optics limit corresponds to the impact parameter associated with the photon orbit. By computing the number of field modes inside the photon sphere utilizing a re-expressed form of the Hawking temperature in terms of the ``charge spring constant'' we introduced, we then employ the brick wall model to compute the smallest distance from the black hole a Hawking particle can be emitted from. This cannot happen arbitrarily close to the horizon, for otherwise, counting field modes is invalid. However, for all ``practical'' purposes this can be treated as the stretched horizon, just barely a Planck length away from the event horizon.

\begin{acknowledgments}
YCO thanks the National Natural Science Foundation of China (No.11705162, No.11922508) and the Natural Science Foundation of Jiangsu Province (No.BK20170479) for funding support.  FG acknowledges funding from state-targeted program ``Center of Excellence for Fundamental and Applied Physics" (BR05236454) by the Ministry of Education and Science of the Republic of Kazakhstan. MG is also funded by the ORAU FY2018-SGP-1-STMM Faculty Development Competitive Research Grant No. 090118FD5350 at Nazarbayev University. 
\end{acknowledgments}

\appendix
\renewcommand{\thesection}{A}

\section{Scaling for a Thermal Ball of Radiation}\label{ball}
Black holes have so much more entropy compared to ordinary matter of the same mass because the scaling is 
\be S_{BH} \sim M^2, \ee
while a thermal ball of radiation scales as
\be S_{R} \sim M^{3/2}. \label{thermalballS}\ee
For the same reason the wavelength of the typical emitted particle will also scale differently.  This is because for the ball of thermal radiation, which could be the source of black hole formation (such as a star prior to gravitational collapse), the Stefan-Boltzmann law gives volume times fourth power of temperature,
\be M \sim T^4 R^3, \ee
We then know that $R\sim M$, is the size of the ball to form a black hole, so that
\be M \sim T^4 M^3 \Longrightarrow M^{-2} \sim T^4, \ee
or just, rearranging for temperature,
\be T\sim M^{-1/2}, \label{balltemp}\ee
so when one takes the derivative of the Stefan-Boltzmann law, one sees that entropy scales as 
\be S\sim T^3 R^3, \ee
which gives, upon plugging in Eq.~(\ref{balltemp}), and $R\sim M$,
\be S\sim M^{3/2}, \ee
which is Eq.~(\ref{thermalballS}), the entropy of a thermal ball of radiation. Thus, the wavelength for a radiation ball will scale as
\be \lambda \sim M^{1/2},\ee
rather than 
\be \lambda \sim M, \ee
as for a black hole.

\renewcommand{\thesection}{B}

\section{A Remark on Local Temperature}\label{nearhor}

One should ask whether $\lambda_{T_H}$ is a physically meaningful scale of the problem.
In flat space, for a fixed temperature the wavelength $\lambda$ is constant, i.e., the value of $\lambda$ at infinity can be directly compared to the size of a body of radius $R$ far away. However, in curved space, the wavelength undergoes redshift as it travels up the gravitational well. The Hawking temperature $T_H$ is the temperature measured by asymptotic observers, why should its associated wavelength be compared directly to the size of the black hole ``infinitely far away'', as in Eq.(\ref{w})? A better ``local'' question to ask would be:
\begin{quote}
\emph{At what ``distance'' $r=\zeta r_\text{h}$ away from the black hole should the Hawking particle be emitted, so that its wavelength is $\mathcal{O}(r_h)?$}
\end{quote}
To answer this question, we shall consider the local temperature given by the Tolman's expression (seen by a stationary observer at coordinate distance $r$):
\begin{equation}\label{tolman}
T_{\text{local}} = \frac{T_H}{\sqrt{1-\frac{2M}{r}}}.
\end{equation}
For explicitness, let its wavelength be $\lambda_{\text{local}} = r_{h}$. Then we want to solve for the multiple $\zeta$ in the equation:
\begin{equation}\label{r2}
\lambda_{\text{local}} = \frac{2\pi}{T_{\text{local}}} = r_{h}.
\end{equation}
With $r= \zeta r_h$, we have
\begin{equation}
8\pi^2r_{h}\left(1-\frac{r_{h}}{\zeta r_{h}}\right)^{\frac{1}{2}}=r_{h}.
\end{equation}
This yields 
\begin{equation}
\zeta = \left[1-\frac{1}{(8\pi^2)^2}\right]^{-1} \approx 1.00016.
\end{equation}

This means that even a Hawking particle emitted ``near'' the horizon has wavelength $\mathcal{O}(r_{h})$. Thus this still agrees with Giddings' remark that Schwazschild black hole does not behave like a typical black body, whose radiation has wavelength much smaller than the size of the body. So even if we use the local temperature, the qualitative picture does not change by much. Of course, in the near horizon limit $\zeta = 1+ \varepsilon$, the wavelength goes like $\sim 8 \pi^2\sqrt{\varepsilon}r_{h}$, so that for a Hawking particle that is emitted \emph{very close} to the horizon, it has very small wavelength (this is just the ``infinite blueshift'' that one might expect). \newline 

With the local temperature, its associated wavelength $\lambda_\text{local}$ satisfies
\begin{equation}
\frac{\lambda_{T_\text{local}}}{r_{h}} = {8\pi^2}\sqrt{1-\frac{r_{h}}{r}},
\end{equation}
c.f. Eq.(\ref{r2}).

\renewcommand{\thesection}{C}

\section{Temperature of Reissner-Nordstr\"om Black Hole Re-Expressed}\label{RNtemp}
The usual expression for the temperature of a charged black hole:
        \begin{equation} T(Q,M)= \frac{\kappa_{RN}}{2\pi} = \frac{\sqrt{M^2-Q^2}}{2\pi \left(\sqrt{M^2-Q^2}+M\right)^2}, \end{equation}
in analogy to the Kerr case \cite{1412.5432}, can be re-expressed by ``peeling'' off the uncharged surface gravity piece, $g \equiv 1/(4M)$:
    \begin{equation} 2\pi T = g - M\Omega^2,\end{equation}
    where $\Omega$ is the ``frequency'',
    \begin{equation} \Omega \equiv \frac{1}{r_+} - \frac{1}{r_s}. \end{equation}
  Here $r_s = 2M$, the uncharged Schwarzschild radius, and $r_+ = M+\sqrt{M^2-Q^2}$, the smaller ($r_s/2<r_+<r_s$) charged outer radius.  So one can see that the spring analogy introduced in \cite{1412.5432} holds in the charged case: $ 2\pi T = g - k_Q$, 
 where $k_Q \equiv M \Omega^2$, suggesting that $\Omega$ holds important physical status as a characteristic frequency for the RN solution, in the same way that in the Kerr case, $\Omega_+$, holds important characterization as the ``angular velocity'' of the outer event horizon. 
 
 A straightforward way to derive this is to consider that the first law of black hole mechanics relates the two necessary parameters, $(M,Q)$, the mass and charge of a Reissner Nordstr\"om black hole:
 \begin{equation}
 \d M = \frac{\kappa}{8\pi} \d A + \Phi \d Q,
 \end{equation}
 where $A$ is the outer horizon area, $\kappa$ is the outer surface gravity, $\Phi$ is the outer potential, and $Q$ is the charge.  
 The area is given by
 \begin{equation}
 A = 4 \pi r_+^2 
 \end{equation}
 where $r_+ = M + \sqrt{M^2 - Q^2}$.
 Equivalently, this area is related to other black hole physical parameters by
 \begin{equation}
 M^2 = \frac{A}{16\pi} + \frac{Q^2}{2} + \frac{\pi Q^4}{A}.
 \end{equation}
 Therefore, we can find the surface gravity via the first law, holding the charge fixed,
 \begin{equation}
 \kappa = 8\pi \left.\frac{\partial M}{\partial A}\right|_Q = \frac{1}{4M} - M \left(\frac{2\pi Q^2}{M A}\right)^2.
 \end{equation}
 This is the form we are looking for, i.e. ``peeling'' off the non-rotating surface gravity.  Now since, 
 \begin{equation}
 \Phi = \left.\frac{\partial M}{\partial Q}\right|_A = \frac{Q}{2M} + \frac{2\pi Q^3}{M A},
 \end{equation}
 we can rearrange and have $\Phi/Q - 1/r_s = 2\pi Q^2/(MA)$ where $r_s = 2M$. Therefore we have $ \kappa = g - k_Q$ where $g = 1/(4M)$ and $k_Q = M \Omega^2$ is the Reissner-Nordst\"om version of the ``spring constant'', analogous to the Kerr case we defined in \cite{1412.5432}.

 %%%%%%%%%%%%%%%%%%%%%%%%%%%%

\end{document}